\documentclass[aps,prd,reprint,groupedaddress]{revtex4-1}

\usepackage{amsmath,amsfonts,amssymb,amscd,amsthm,xspace}
\usepackage{graphicx}
\usepackage{epstopdf}
\usepackage{enumerate}
\usepackage{wrapfig}
\usepackage{color}
\usepackage{longtable}
\usepackage{float}
\usepackage{hyperref}
\usepackage{mathtools}
\usepackage{natbib}
\usepackage{xcolor}

\usepackage{times}

\begin{document}


\title{Charged fluid structures around a rotating compact object with a magnetic dipole field}


\author{Kris Schroven}
\email[]{kris.schroven@zarm.uni-bremen.de}
\author{Audrey Trova}
\email[]{audrey.trova@zarm.uni-bremen.de}
\author{Eva Hackmann}
\email[]{eva.hackmann@zarm.uni-bremen.de}
\author{Claus L\"ammerzahl}
\email[]{claus.laemmerzahl@zarm.uni-bremen.de}
\affiliation{University of Bremen, Center of Applied Space Technology and Microgravity (ZARM), 28359 Bremen}


\date{\today}

\begin{abstract}
 We study stationary, electrically charged fluid structures encircling a rotating compact object with a dipole magnetic field oriented along the rotation axis. This situation is described in an idealized way by the Kerr metric and a magnetic dipole "test" field, that does not affect the spacetime. The self-gravitational and self-electromagnetic field of the fluid are neglected and the fluid is assumed to be non conductive and in rigid motion. Our work generalizes a previous study by \citet{kovar16} by taking into account the rotation of the central object. Therefore, we focus on the influence of the rotation onto the existence and position of bound fluid structures. Frame dragging effects allow the existence of polar clouds, which could not be found in non-rotating case. Furthermore counter-rotating equatorial tori become more preferred the faster the central object is spinning.
\end{abstract}


\maketitle


\section{Introduction}
	Fluids take a very important role in Astrophysics. Accreted by compact objects such as black holes or neutron stars, they give rise to a variety of astrophysical phenomenon like active galactic nuclei (AGN), X-ray binaries and more \cite{frank,yuan}. Their investigation however is a very challenging issue. The density, pressure and temperature of the fluid lies in a very broad range, so that different approaches are needed to describe different situations, where, according to the situation, we  have to include radiation processes, turbulences, nuclear burning electromagnetic interactions and more. For very diluted fluids particles do not interact and are described by the test particle approach \cite{prendergast,kovar08}. A kinetic description is used for less diluted fluids (see \cite{buchplasma} and citations within) whereas the magnetohydrodynamic (MHD) description is suitable for dense fluids \cite{rezzolla13,punsly8}.  Drastic simplifications of the full picture are therefore needed to build (analytic) models of accretion discs, like the thin disc model, the slim disc model, ADAFs, Polish Doughnuts and more (see \cite{Abramowicz2013} and citations within). These models play a very important role in understanding the general physical processes in accretion discs. They are also used to simplify numerical simulations or serve as initial conditions or test beds to the simulations.
	
	Thick accretion discs with a negligible loss of mass can be modeled analytically in a general relativistic background with the Polish Doughnut model, that uses a hydrodynamical, perfect fluid description for the fluid. In this model gravity plays a crucial role for building toroidal configurations. The model was introduced 1978 for a neutral fluid by \cite{pd78} in the case of a Schwarzschild background, then studied for Kerr \cite{pd78kerr}, and later on for more complicated backgrounds \cite{TDdesitter,paul16}.
	
	Magnetic fields are present during most accretion processes, produced either by the accreted fluid itself, by the object accreting the matter (e.g.~a magnetar) or as an external magnetic field (eg. an interstellar one). These fields will have a major effect on the accretion of plasma, or on an otherwise charged fluid (e.g.~a dusty fluid charged by its interaction with the energetic radiation from an AGN \cite{dustyplasma}). In regards to that the Polish Doughnut model was extended. A toroidal magnetic field produced by the fluid was added to the model in \cite{komissarov06}, while in \cite{kovar14,kovar16, trovar18} the interaction of a charged fluid with an external magnetic field was considered.
	
	In this work we build up on the results in \cite{kovar16} and investigate charged perfect fluids encircling compact objects while located in a electromagnetic background field. The charged fluid takes on structures, that are constructed within a model derived from the conservation laws and Maxwell equations as well as the usual assumptions of stationarity and axial symmetry in the Polish Doughnut model. Self-gravitational and self-electromagnetic fields of the fluid configuration as well as the influence of the electromagnetic background field on the spacetime are neglected in our setting. A charge distribution has to be assigned to the fluid, that is approximated as fully non conductive - the opposite approximation to the infinite conductivity assumed in the ideal MHD approach to plasma description. The angular momentum profile and equation of state of the fluid configuration are chosen beforehand, so that its pressure and energy density profile can be calculated.

	The fluid encircles the compact object with a constant angular velocity, which corresponds to an angular momentum profile with an increasing angular momentum for bigger radii. This assumption has the benefit that the problem can be solved analytically. It was shown for the uncharged case, that structures with a constant angular momentum show a runaway instability \cite{abramowicz1983}, which leads to an almost complete accretion of the torus by the central object on dynamical time scales. This instability is suppressed for an increasing angular momentum profile towards bigger radii \cite{daigne2004}. This behavior in the uncharged case gives some motivation to the assumptions of a rigidly rotating fluid. While equilibrium tori in rigid rotation are impossible for the uncharged case, we show that this is not a problem for charged fluids. The fluid is described by a polytropic equation of state. Fluid configurations might form bound structures anywhere around the compact object. We will, however, focus our study on fluid configurations, which centers lie either in the equatorial plane (called equatorial tori), or on the rotation axis (referred to as polar clouds). 
	
	After introducing a general procedure to look for possible fluid structures, we specify to the special case of a Kerr metric and a dipole magnetic field, that is oriented along the symmetry axis in the Kerr metric. This combination of metric and field describes in an idealized way a rotating compact object, that produces a magnetic dipole field (eg. a magnetar), while the non-conductive fluid might describe partly ionized helium.
	Since this set up was already discussed for the Schwarzschild metric by \citet{kovar16}, our main interest concerns the influence of the rotation of the central object on the shape and existence of the fluid structures. 
	
	This paper is organized as follows. In Sec. \ref{theory} the model for the construction of charged fluid configurations is described. The main pressure equations are derived from the conservation laws and Maxwell equations and solved in terms of an effective potential. Equations for the physical characteristics pressure, energy density and charge distribution of the fluid are given. The general procedure of how to find possible bound fluid structures is presented in Sec. \ref{procedure}. In Sec. \ref{kerrcase} we specify to the case of a Kerr metric and a dipole magnetic field. The Kerr metric and the electromagnetic potential for a dipole magnetic field in Kerr are given and shortly discussed. We take a look onto the uncharged limit and discuss the behavior of the effective potential of the fluid structures in the charged case. The behavior of solutions for equatorial tori and polar clouds in regards to various parameters are discussed in Sec. \ref{equatorialtori} and \ref{polarclouds} respectively. For both cases examples for a fluid structure and its physical characteristics are given. Conclusions are given in Sec. \ref{conclusion}.
	
	Throughout the paper the geometrical system of units ($c=G=k_B=1$) is used. In case that the physical (SI) units are used the quantities are indicated by the index SI. 

\section{Charged fluid structures in an external electromagnetic test field \label{theory}}
 \subsection{General assumptions of the thick disc model}\label{sec:assumptions}
 To build a charged fluid torus located in an external electromagnetic test field, we will follow the approach made in \cite{kovar14} and use the general setup for Polish Doughnuts \cite{rezzolla13,Abramowicz2013}. Therefore we make the following assumptions:
 
  1) The fluid, which builds the accretion disc, has a negligible effect on the spacetime metric. It therefore serves as a "test-fluid", positioned in a given background spacetime. The fluid is furthermore described as a perfect fluid with a polytropic equation of state.
  
  2) The considered spacetime is axially symmetric and stationary. In Boyer-Lindquist coordinates the metric takes the form
\begin{align}
ds^2=g_{tt}\,dt^2+2g_{t\phi}\,dt\,d\phi+g_{rr}\,dr^2
    +g_{\theta\theta}\,d\theta^2+g_{\phi\phi}\,d\phi^2.
\end{align}
  It is required that the electromagnetic test field is stationary and axially symmetric as well. This implies that in a certain gauge the electromagnetic vector potential has the form 
  \begin{align}
 A_{\mu}=(A_t,A_\phi,0,0)  \,.
  \end{align}
  
  3) The fluid is also axially symmetric and stationary, with purely circular motion. The four velocity for that case can be written as 
 \begin{align}
  U^\mu=\left(U^t,U^\phi,0,0\right).
 \end{align}
  Specific angular momentum and angular velocity are defined as
 \begin{align}
  \ell=-\frac{U_\phi}{U_t},~~~~~~~\omega=\frac{U^\phi}{U^t},
   \label{lomega}
 \end{align}
  and are connected by the relation
  \begin{align}
  \omega=-\frac{\ell\,g_{tt}+g_{t\phi}}{\ell\,g_{t\phi}+g_{\phi\phi}}\,.
  \end{align}
  Finally the $t$-component of the four-velocity can be derived by using the normalization condition, and takes the form
  \begin{align}
   (U^t)^2=-\frac{1}{g_{tt}+\omega g_{t\phi} + \omega^2 g_{\phi\phi}}\,.
   \end{align}  
 
 \subsection{Pressure equations for a charged fluid}
 The pressure equations in a thick disc model can now be derived by solving the conversation law
 \begin{align}
 \nabla_\nu T^{\mu\nu}=0\,,
 \label{conslaw}
 \end{align}
 where $(T^{\mu\nu})$ is the energy momentum tensor and $\nabla_\nu$ indicatest the covariant derivative. In case of a charged fluid tori, it can be split into two terms, a matter term ($T^{\mu\nu}_{MAT}$) and an electromagnetic term ($T^{\mu\nu}_{EM}$),
 \begin{align}
 T^{\mu\nu}_{MAT} & =(\epsilon+p)U^\mu U^\nu + p\, g^{\mu\nu}\, ,\\
 T^{\mu\nu}_{EM} & =\frac{1}{4\pi}\left({F^\mu}_\gamma F^{\nu\gamma}-\frac{1}{4}F_{\gamma\delta} F^{\gamma\delta}\, g^{\mu\nu} \right)\, ,
 \end{align}
 where $ F_{\mu\nu}=\nabla_\mu A_\nu-\nabla_\nu A_\mu=\partial_\mu A_\nu-\partial_\nu A_\mu  $ is the electromagnetic tensor, $A_\mu$ the axially symmetric and stationary total electromagnetic potential, and $\epsilon$ and $p$ denote the energy density and pressure of the fluid.
 
 In this case the Maxwell equations 
 \begin{align}
 \nabla_\nu F^{\mu\nu}=&4\pi\, J^\mu\, , \\
 \nabla_{\left(\gamma\right.}F_{\left.\mu\nu\right)}	=&0
 \end{align}
 have to be considered as well to derive the pressure equations. Here $J^\mu=\rho_q\, U^\mu+\sigma F^{\mu\nu}U_\nu$ is the four-current, with the conductivity $\sigma$ and charge density $\rho_q$. 
 By assuming that the internal electromagnetic field produced by the charged fluid is much smaller than the external test field ($F^{\mu\nu}_{INT}\ll F^{\mu\nu}_{EXT}$), and by further demanding that the conductivity vanishes ($\sigma=0$), we derive \cite{misner1973}
  \begin{align}
 \nabla_\nu T^{\mu\nu}_{EM}=-F^{\mu\nu}_{EXT}\,J_\nu  \quad \text{  with } J^\mu=\rho_q\, U^\mu\, .
 \end{align}
 By using Eqn. \eqref{conslaw}, this leads to the following main equation
 \begin{align}
 \nabla_\nu T^{\mu\nu}_{MAT}=F^{\mu\nu}_{EXT} J_\nu \, .
 \label{masterformular}
 \end{align}
 In case of a neutral fluid, Eqn. \eqref{masterformular} reduces to Eqn. \eqref{conslaw}, where $T^{\mu\nu}$ then contains the matter term only. Note that the assumption of zero conductivity is a necessary condition for the self-consistency of the model. A non-zero conductivity would allow radial electric currents. This is, however, in contradiction to the thick disc model, where a circular motion of the fluid is required. 
 
 The pressure equations for a charged thick disc located in an external electromagnetic test field now follow directly from Eqn. \eqref{masterformular} \cite{kovar14}
 \begin{align}
 \partial_\mu p=&(p+\epsilon)\left(\partial_\mu \ln \left( U^t\right) -\frac{\ell\partial_\mu \omega}{1-\omega\ell}\right. \nonumber\\
                &\left.+\frac{\rho_q}{p+\epsilon}\left( U^t\partial_\mu A_t +U^\phi\partial_\mu A_\phi\right)\right)\, .
 \label{peq}
 \end{align}
 The electromagnetic force on the charged fluid in $\mu$-direction is 
 \begin{align}
	\left.\begin{matrix} \text{repulsive for } 0< \\ \text{attractive for } 0> \end{matrix} \right\} \frac{\rho_q}{p+\epsilon}\left( U^t\partial_\mu A_t+U^\phi \partial_\mu  A_\phi \right)\, .
	\label{forcedir}
 \end{align}
 
 Because of the discussed symmetries of the model Eqn. \eqref{peq} is different from zero only for $\mu=\theta ,r$, leading to two pressure equations.
 
 \subsection{Integrability condition - restrictions to the charge distribution}
 For the pressure Eqns. \eqref{peq} to be solvable, the integrability condition 
 \begin{align}
 \partial_\mu\left(\partial_\nu p\right)=\partial_\nu\left(\partial_\mu p\right)
 \label{intcon}
 \end{align}
 has to be satisfied. In case of an uncharged fluid, where the last term in Eqn. \eqref{peq} vanishes, Eqn. \eqref{intcon} is fulfilled, if the fluid has a barotropic equation of state $\epsilon=\epsilon(p)$ (see eg. \cite{Abramowicz2013}). By keeping the assumption of a barotropic equation of state and by setting
  \begin{align}
   \mathcal{K}=\frac{\rho_q}{p+\epsilon} \label{kkk}
 \end{align}
 the last term in Eqn. \eqref{peq} has to satisfy 
  \begin{align}
 \partial_\mu [\mathcal{K}\,U^t\left(\partial_\nu A_t +\omega\partial_\nu A_\phi\right) ] = \partial_\nu [\mathcal{K}\,U^t\left(\partial_\mu A_t +\omega\partial_\mu A_\phi\right) ] \label{intcon2}
 \end{align}
 to fulfill the integrability condition \eqref{intcon}. For Eqn. \eqref{intcon2} to hold we have to specify some additional constraint on the charge distribution $\rho_q$ contained in $\mathcal{K}$, and/or the radial distribution of the angular momentum $\ell$, which is related to $\omega$ (see Eqn. \eqref{lomega}). 
 
 We will here restrict our model further, by assuming the charged fluid to be in a rigid rotation, and set $\omega$ to be constant. Equation \eqref{intcon2} can then be written as
 \begin{align}
 \partial_\mu [\mathcal{K}\,U^t\partial_\nu\left( A_t +\omega A_\phi\right) ] = \partial_\nu [\mathcal{K}\,U^t\partial_\mu\left( A_t +\omega A_\phi\right) ]\, .
 \label{intconwc}
 \end{align}
 In analogy to demanding a barotropic equation of state to make the first term of \eqref{peq} satisfy the integrability condition, we can now easily fulfill the integrability condition \eqref{intconwc} for the second term by assuming 
 \begin{align}
 \mathcal{K}\,U^t=f_\mathcal{K}\left(S\right), \text{ for } S=A_t+\omega A_\phi\, ,
 \label{kdef}
 \end{align} 
 where $f_\mathcal{K}(S)$ is an arbitrary function of $S$. Here $S$ corresponds to an electromagnetic potential acting on a charged particle with an angular velocity $\omega$ on a circular orbit. Curves of a constant $S$ will coincide with curves of constant $f_\mathcal{K}(S)$.
 
 Note that if one of the two components $A_t$, $A_\phi$ vanishes, the assumption of rigid rotation of the fluid is not necessary anymore and the condition for the charge distribution reduces to $\mathcal{K}U^\alpha=f_\mathcal{K}(A_\alpha)$.
 
 \subsection{Solutions of the pressure equations}
 The pressure Eqns. \eqref{peq} can be rewritten in terms of an effective potential $h$ defined by
  \begin{align}
 h=\frac{\Gamma-1}{\Gamma}\,\left(\int^p_0{\frac{dp}{p+\epsilon}}\right)\,.
 \label{inth}
 \end{align}
 Using
 \begin{align}
 \label{Pdef}
 \mathcal{P}=\frac{1}{{(U^t)}^2}= -\left( g_{tt}+\omega g_{t\phi} + \omega^2 g_{\phi\phi}\right)\, ,
 \end{align}
Eqns. \eqref{peq} read
 \begin{align}
 \partial_\mu h(r,\theta)= \frac{\Gamma-1}{\Gamma}\,\left(-\frac{\partial_\mu \mathcal{P}}{2 \mathcal{P}}+ f_\mathcal{K}(S)\,\partial_\mu S\right)\, ,
 \label{dh}
 \end{align}
 where we used the assumption of rigid rotation ($\partial_\mu\omega=0$), and the condition for the charge distribution in Eqn. \eqref{kdef}. The prefactor $\frac{\Gamma-1}{\Gamma}$ is necessary later on to describe the physical characteristics in terms of $h$ in a nice way. Integrating Eqn. \eqref{dh} leads to the following effective potential $h$,
 \begin{align}
 h=\frac{\Gamma-1}{\Gamma}\,\left(-\frac{1}{2}\ln{\mathcal{P}}+\int{f_\mathcal{K}(S) d S}\right)+h_0\,.
 \label{potrigid}
 \end{align}
 Here $h_0$ is an integration constant. It allows to choose the point $(r,\theta)$, where the effective potential becomes zero.
 
 If an explicit equation for $\epsilon(p)$ is given, the pressure $p$, energy density $\epsilon$ and charge density $\rho_q$ of the fluid can be expressed in terms of $h$.  As can be seen from Eqn. \eqref{inth}, equipotential surfaces of $h$ coincide with surfaces of constant pressure $p$, and therefore of constant $\epsilon$ as well.
 
 A bound solution for a stationary charged fluid structure (e.g. an equatorial torus) in the given setup exists if we can find a local maximum for the effective potential $h$ at a position $(r_c,\theta_c)$. The point $(r_c,\theta_c)$ then defines the center of the structure, and the outer edge is given by $h=0$, where also the pressure vanishes (see Eqn. \eqref{inth}). Necessary existence conditions for fluid structures are therefore given by 
 \begin{align}
\partial_{\theta}h(r_c,\theta_c)&=0\,, & \partial_{r}h(r_c,\theta_c) &=0\,.
 \label{excon1}
 \end{align}
 Furthermore, to guarantee that the local extrema is indeed a maximum, the Hessian matrix for $h$ 
 \begin{align}
 \mathcal{H}=\left(\begin{matrix}
 \partial^2_{rr}h & \partial^2_{r\theta}h \\
 \partial^2_{\theta r}h & \partial^2_{\theta\theta}h
 \end{matrix}\right)			
 \end{align} 
 has to be negative definite at the point $(r_c,\theta_c)$. This gives the sufficient conditions
 \begin{align} 
\partial^2_{rr}h(r_c,\theta_c) < 0 \quad \text{and} \quad \det(\mathcal{H})(r_c,\theta_c) > 0\,.
\label{excon1.2}
 \end{align}
 
 Saddle points $(r_s,\theta_s)$ can behave as so called cusp points, when they occur additionally to the maximum in a solution for the fluid structure. If the integration constant $h_0$ is chosen such that $h=0$ at the saddle point, it might serve as a point where fluid material can flow out of the structure (e.g. out of the thick disc). However, since several saddle points can occur at various positions and $h$-values for some solutions, not every saddle point will actually behave as a cusp point for a given structure.  
 
 \subsection{Physical characteristics}
 The choice of the equation of state determines the connection between pressure and energy density.
 Following \citet{kovar14} and \citet{trovar18} we choose a polytropic equation of state for the fluid, 
 \begin{align}
 p=\kappa\epsilon^\Gamma\, ,
 \label{eos}
 \end{align} 	  
 where $\kappa$ and $\Gamma$ are the polytropic coefficient and exponent respectively.
 
 The pressure $p$, energy density $\epsilon$ and specific charge density $q=\frac{\rho_q}{\epsilon}$ are then given in terms of the effective potential $h$. By plugging Eqn. \eqref{eos} into Eqn. \eqref{inth} we find
 \begin{align}
 p=\left(\frac{e^{h}-1}{\kappa^{\frac{1}{\Gamma}}}\right)^{\frac{\Gamma}{\Gamma-1}}\, ,\\
 \epsilon=\left(\frac{e^{h}-1}{\kappa}\right)^{\frac{1}{\Gamma-1}}\, .
 \end{align}
 By further defining the specific charge density $q=\frac{\rho_q}{\epsilon}$ and using \eqref{kkk} we get
 \begin{align}
   q=\frac{\rho_q}{\epsilon}=\mathcal{K}\,e^{h}\, .
 \end{align}
 The total mass and charge of the charged fluid structure can be calculated by integrating the mass density $\rho$ and the charge density $\rho_q$ over the whole volume $\mathcal{V}$ of the structure,
 \begin{align}
 \mathcal{M}=\int_\mathcal{V} \rho\, d\mathcal{V}\, , \\
 \mathcal{Q}=\int_\mathcal{V} \rho_q\, d\mathcal{V}\, .
 \end{align}
 
 Up to now, the introduced setup for charged fluid structures in an external magnetic field has not made any statements concerning the mass density $\rho$. However, by introducing a suitable assumption for $\rho$ as $\rho=\rho(\epsilon,p)$, the mass density can also be derived from the effective potential $h$. In the non-relativistic limit an appropriate assumption would be $\rho\approx\epsilon$.
 
 	Following the approach in \cite{kovar14}, the magnetic field strength of the fluid torus $\mathcal{B}$ is approximated at the edge $r_{out}$ of the torus by  a charged ring that contains the charge of the whole torus $\mathcal{Q}$ and rotates at the same angular velocity $\omega$,
 	\begin{align}
 	\mathcal{B}\approx \frac{\omega \mathcal{Q}}{\pi (r_{out}-r_c)}\, .
 	\label{bapprox}
 	\end{align}
The total mass and charge of a fluid structure as well as its magnetic field strength have to be sufficiently low to not violate the assumptions of our model. If the effective potential $h$ for a solution is found, these requirements restrict the possible choices for $h_0$ or the polytropic coefficient $\kappa$.

 	Here the magnetic field strength $\mathcal{B}$, the dipole moment $B$, the charge $\mathcal{Q}$, the angular velocity $\omega$ and the radius $r$ are all given in dimensionless units. They can be transfered back into $SI$ units by
 	\begin{align}
 	\mathcal{Q}_{\rm SI} & = Mc^2\sqrt{\frac{4\pi\epsilon_0}{G}}\, \mathcal{Q}\,, & B_{\rm SI}=&\frac{c}{M\sqrt{4\pi\epsilon_0 G}}\,B\,, \nonumber\\
 	\omega_{\rm SI} & = \frac{c}{M}\, \omega\,, & \mathcal{B}_{\rm SI}=&\frac{c}{M\sqrt{4\pi\epsilon_0 G}}\,\mathcal{B}\,,\\
 	r_{\rm SI} & = Mr\, . \nonumber
 	\end{align}
 	Here $G$ is the gravitational constant, $\epsilon_c$ the electric constant and $M=\frac{G m}{c^2}$, where $m$ is the mass of the central object.
 
 \section{Construction of charged fluid structures \label{procedure}}
 If the spacetime metric and the electromagnetic potential of the external test field show, next to the required axial symmetry and stationarity, a mirror symmetry at $\theta=\frac{\pi}{2}$, and are furthermore differentiable at $\theta=0$, than the following procedure can be used to find charged fluid structures in the given setting.
 
 The first step is to make sure, that the existence conditions \eqref{excon1} are fulfilled for a given position $(r_c,\theta_c)$, where the structure's center is located. Due to the required symmetries and assumptions the first condition in \eqref{excon1} will always be fulfilled for $\theta_c=0,\frac{\pi}{2},\pi$.
 We will therefore focus our search for fluid structures to
 \begin{itemize}
 	\item tori centered in the equatorial plane ($\theta_c=\frac{\pi}{2}$), and
 	\item polar clouds centered on the polar axis ($\theta_c=0,\pi$).
 \end{itemize}
 The second condition in \eqref{excon1} can be used as a normalization condition for the function $f_\mathcal{K}(S)$ connected to the charge distribution. From Eqn. \eqref{dh} we get the following relation that has to hold at the center of the structure,
 \begin{align}
f_\mathcal{K}(S)(r_c,\theta_c) = \frac{\partial_r \mathcal{P}}{2 \mathcal{P}} \frac{1}{\partial_r S}\Big|_{r=r_c,\theta=\theta_c} \eqqcolon  b\,.
 \label{fcenter}
 \end{align}
 If $g(S)$ is an arbitrary normalized function of $S$, meaning $g(S)=1$ at the point $(r_c,\theta_c)$, then we can choose $f_\mathcal{K}(S)$ as follows so that the second existence condition in \eqref{excon1} is always satisfied,
 \begin{align}
f_\mathcal{K}(S) = b\, g(S)\,.
 \label{fdef}
 \end{align}
 
 In a second step one has to make sure, that the conditions for a local maximum \eqref{excon1.2} are fulfilled. In case of an electromagnetic potential $A^\mu$ and a spacetime metric with a mirror symmetry at $\theta=\pi/2$ (and the metric and electromagnetic potential being differentiable), the mixed partial derivatives of $h$ vanish at $\theta=0,\frac{\pi}{2}$, and the conditions for the maximum reduce to
 \begin{align}
 \partial^2_{rr}h (r_c,\theta_c)<0,~~\partial^2_{\theta\theta}h(r_c,\theta_c)<0\, .
 \label{excon2}		
 \end{align}
 By using the result \eqref{potrigid} in the sufficient conditions \eqref{excon2} we find 
 \begin{align}
 0>&\frac{\Gamma-1}{\Gamma}\left\{-\mathcal{P}\partial^2_{rr}\mathcal{P}+ \left(\partial_r\mathcal{P}\right)^2\right.\nonumber\\
                                       &\left.\left.+\mathcal{P}\frac{\partial_r \mathcal{P}}{\partial_r S}\, \left(\partial^2_{rr} S+\frac{f'_\mathcal{K}}{f_\mathcal{K}}(S)\, \left(\partial_r S\right)^2\right)\right\}\right|_{\substack{r=r_c\\\theta=\theta_c}}\, ,
 \label{ineqex1}\\
 0>&\left.\frac{\Gamma-1}{\Gamma}\left\{-\partial^2_{\theta\theta}\mathcal{P}+\frac{\partial_r \mathcal{P}}{\partial_r S}\, \partial^2_{\theta\theta} S\right\}\right|_{\substack{r=r_c\\\theta=\theta_c}}\, .
 \label{ineqex2}
 \end{align}
 (Please note that for saddle points the right hand sides in \eqref{ineqex1}-\eqref{ineqex2} are both non-zero, but only one of the two inequalities is satisfied). While \eqref{ineqex1} can be satisfied for arbitrary angular velocities $\omega$ by a proper choice of $f_\mathcal{K}(S)$, the second inequality can be fulfilled by restricting the choice for $\omega$. In the case that is discussed in the following we choose the arbitrary function $g(S)$ in Eqn. \eqref{fdef} explicitly, before satisfying both conditions \eqref{ineqex1}-\eqref{ineqex2} by a restriction to the choice of $\omega$.  By doing so it is possible to compare our results with the work by \citet{kovar16}, which is the Schwarzschild limit to our setup (also defining $g(S)$ first helps to not run into crazy charge distributions).
 
 According to Eqn. \eqref{forcedir}, a repulsive electromagnetic force acts on the charged fluid in the radial direction, if $f_\mathcal{K}(S)\partial_r S$ is bigger than zero, and an attractive force otherwise. At the extrema $r_e$ of $h$ (which include cusp points at $\theta=\pi/2,0$ and the center $r_c$ of the structure), by using the connection given in \eqref{fcenter}, the condition for a repulsive electromagnetic force reduces to 
 \begin{align}
 	\left.\partial_r \mathcal{P}\right|_{\substack{r=r_e\\\theta=0,\frac{\pi}{2}}}>0\,.
 \end{align}
 
 After a local maximum in the effective potential is found, the integration constant $h_0$ is chosen to determine the outer edge of the fluid structure. If the edge of the structure passes through a saddle point of the potential, this might create a cusp point. Finally, one has to make sure, that the initial assumption of a negligible electromagnetic field of the fluid structure is still valid ($F^{\mu\nu}_{INT}<<F^{\mu\nu}_{EXT}$). This can be accomplished by setting the density of the charged fluid sufficiently small by choosing the scaling factor $\kappa$ in the equation of state \eqref{eos} accordingly. Limits to the diluteness of the fluid are given by the magnetohydrodynamic approach, which needs to still be applicable. 
 
 \section{The case of a Kerr metric with a magnetic dipole test field \label{kerrcase}}

	\begin{figure}
 		\includegraphics{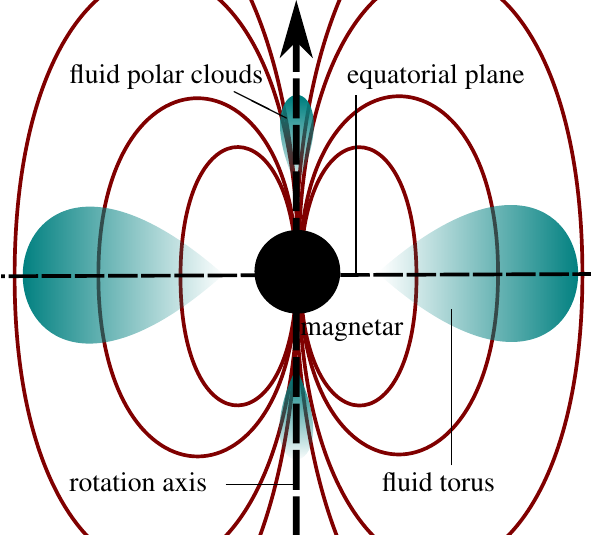} 
 	\caption{Sketch of the considered situation. The dipole magnetic test field is aligned to the rotation of the compact object (e.g. a magnetar). }
 	\label{model}
 	\end{figure}

 In this work we want to discuss possible charged fluid structures around a rotating compact object, that produces a dipole magnetic field. This field is aligned to the rotation axis of the compact object. This situation will be described by a Kerr metric with an external dipole magnetic test field. The results are then compared to the Schwarzschild limit discussed in \cite{kovar16}, and the charged fluid structures found in \cite{trovar18} in case of a Kerr metric with an external homogeneous magnetic field. A sketch of the considered situation is shown in Fig. \ref{model}. Obviously, both the Kerr metric and the potential of the dipole magnetic test field have the required symmetries (axial symmetry, stationarity, and mirror symmetry to $\theta=\frac{\pi}{2}$) for the construction procedure outlined in section \ref{procedure}.
 
   	 To motivate our considered model, we first  want to shortly sum up the discussion in \citet{kovar16}, about which scenario could be described (in a very idealized way) by the given model of a charged, non-conducting fluid circulating in a Kerr (Schwarzschild in \cite{kovar16}) background with a dipole magnetic test field (see \cite{kovar16} for details).
 	
	 The central object of mass $m$, that is mimicked by the Kerr-metric, should be very compact, so that the radius does not exceed $3M$, where $M$ is the Schwarzschild radius given by $M=Gm/c^2$. The object produces the magnetic dipole field, which is considered in our model. A compact rotating neuron star with a strong magnetic dipole field of $B=10^8T$ could be described like that in a very idealized way, especially since we further have to assume that the dipole field has to be oriented along the rotation axis of the neutron star.
 
	 A fluid with a non vanishing charge, but zero-conductivity might describe a partly ionized helium fluid, in case of high pressures and low temperatures, which implies high densities.
 
	 An open question is still, how the given charge distribution within the fluid is reached, which is necessary so that the integrability condition is satisfied, and if the distribution is stable. 
 
 A mathematical description of the Kerr metric and the electromagnetic potential of the dipole magnetic test field will be given in the following before discussing the behavior of the effective potential $h$ at the equatorial plane and the poles, both in the case of a charged fluid as well as the uncharged limit.
 
 \subsection{Kerr metric and the magnetic potential} \label{elmpotdisc}
 The Kerr metric in Boyer-Lindquist coordinates and geometrical units $c=1$, $G=1$ is given by
 \begin{align}
 ds^2=&\frac{\Sigma}{\Delta}dr^2+\Sigma d\theta^2+\frac{\sin^2\left(\theta\right)}{\Sigma}\left[\left( r^2+a^2\right)d\phi -a\;dt\right]^2 \nonumber \\
 &-\frac{\Delta}{\Sigma}\left[a \sin^2\left(\theta\right)d \phi-dt\right]^2~~,
 \end{align}
 where
 \begin{align}
 \Sigma(r,\theta)&=r^2+a^2 \cos^2\left(\theta\right)~~,\\
 \Delta(r)&=r^2-2 r+a^2~~.
 \label{delta}
 \end{align}
Here we further normalized all quantities with respect to the mass $m$ of the central object such that they are dimensionless. Accordingly, $a$ is the normalized angular momentum $0\leq a$. The horizons of a Kerr black hole are given by $\Delta(r)=0$, i.e.~$r_\pm = 1\pm\sqrt{1-a^2}$.
 
 The frame dragging effect in Kerr spacetime connects $\phi$- and $t$- components via cross terms in the metric. This leads to an $A_t$-component in the description of the dipole magnetic field. This term will locally give rise to an electric part in the field. The electromagnetic potential for a dipole magnetic test field in Boyer-Lindquist coordinates is given by  \cite{prasanna78}:
 \begin{align}
 A_t =& -\frac{3}{2} \frac{a B}{\xi^2 \Sigma}\left(-\left(r-\cos^2(\theta)\right)
     +\frac{1}{2\xi}\ln{\frac{r-1+\xi}{r-1-\xi}}\right.\nonumber\\ 
     &\left.\times \left(r(r-1)+(a^2-r)\cos^2(\theta)\right)\right)\, ,\\
 A_\phi=&-\frac{3}{4} \frac{ B \sin^2{\theta}}{\xi^2 \Sigma}\left((r-1)\Sigma+2\,r\,(r+a^2)\right.\nonumber\\
        &\left.-\frac{1}{2\xi}\ln{\frac{r-1+\xi}{r-1-\xi}}\,\left(\chi-4\,r\,a^2\right)\right)\, ,
 \end{align}
 where $\xi=\sqrt{1-a^2}$,
 \begin{align}
  \chi (r,\theta)&=(r^2+a^2)^2-\Delta(r)\,a^2\,\sin^2{\theta}\,,
 \end{align}
 and $B$ is the dipole moment of the external magnetic field. One can show that $A_\phi\geq 0$ holds for all $r>r_+$ and for all $\theta$ if $B>0$, while $A_t$ changes signs depending on $\theta$ ($A_t\leq 0$ for $r> r_+$, $\theta=\pi/2$; $A_t\geq 0$ for $r>r_+$, $\theta=0$.).
 
 In the case $a=0$, the electromagnetic potential reduces to the magnetic dipole test field in Schwarzschild spacetime \cite{dpss},
 \begin{align}
 A_t &= 0\,,\\
 A_\phi&=-\frac{3}{4} B \sin^2{\theta}\left(r+1-\frac{r^2}{2}\ln{\frac{r}{r-2}}\right)\,.
 \end{align}		
 Here the potential only contains a $\phi$-component and the electric component the field vanishes.
 
 In the extremal Kerr case ($a=1$), the electromagnetic potential reduces to \cite{Koyama09}
 \begin{align}
 A_t &= -\frac{B}{2(r-1)^2\Sigma}\left(r\sin^2(\theta)-2\,(r-1)\cos^2(\theta)\right) \,,\\
 A_\phi&=-\frac{B\sin^2(\theta)}{2(r-1)^2\Sigma}\left((r-1)(r+\cos^2(\theta))-2\,r^3\right) \,.
 \end{align}
 The electromagnetic potential components fall off and approach zero for big values of $r$, while $A_t$ approaches zero faster than $A_\phi$. For $\theta \neq 0$, the components diverge at the outer horizon to $\pm \infty$.
 
 \subsection{Uncharged limit \label{unchargedlimit}}
 Before discussing the general case it gives some insight to have a look on the limit where the charge of the fluid or the $B$-field vanishes. Solutions for tori or polar clouds exist if $h$ has a local maximum at $\theta_c=0,\pi/2$ respectively. We will show in the following that for the uncharged case no equilibrium structures in rigid rotation can be found.
 
 \begin{figure}
  	\includegraphics{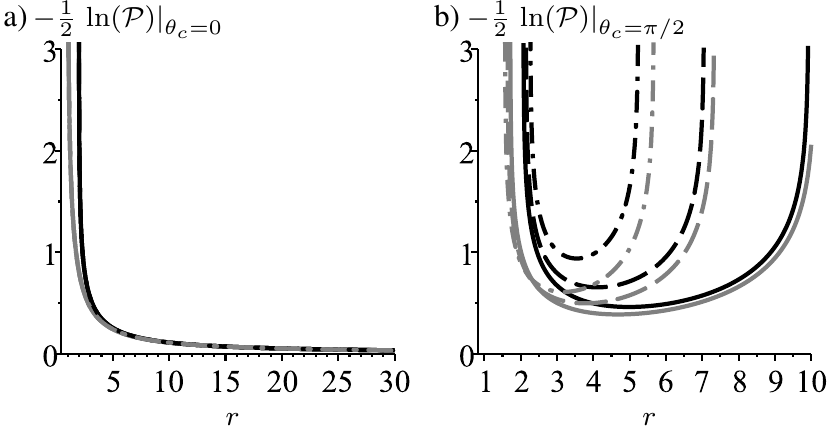} 
 	\caption{$-\frac{1}{2}\left.\mathcal{P}\right|_{\theta=\theta_c}$ plotted over $r$ for (a) $\theta_c=0$ and (b) $\theta_c=\pi/2$. Gray and black lines correspond to $a=1$ and $a=0$ respectively. Solid lines correspond to $\omega=0.09$, dashed lines correspond to $\omega=0.12$ and dash-dotted lines correspond to $\omega=0.15$. The region for $r$ in which the term has real values shrinks with rising values of $\omega$, since superluminal motion of the fluid is reached faster. }
 	\label{pplot}
 \end{figure}

 In the uncharged case the effective potential $h$ given in \eqref{potrigid} reduces to $h|_{q=0}=\frac{\Gamma-1}{\Gamma}(-\frac{1}{2}\ln\mathcal{P}) =: \frac{\Gamma-1}{\Gamma} T_1$. The behavior of $T_1$ is shown in figure \ref{pplot} for $\theta=\pi/2$ and $\theta=0$. It is immediately clear that $h$ (for both $q=0$ and $q\neq 0$) is only defined for $\mathcal{P}>0$ and diverges to $+\infty$ at $\mathcal{P}=0$, where the fluid would reach luminal motion. 
 
 Let us first discuss the case of polar clouds ($\theta=0$). Then $\mathcal{P}$ is given by
 \begin{align}
 	\left.\mathcal{P}\right|_{\theta=0} & =\frac{\Delta(r)}{r^2+a^2}\,,
 	\label{peqn}
 \end{align}	
 which is independent from $\omega$ and approaches one in the limit $r\rightarrow \infty$. Then it is clear that in the uncharged case the effective potential $h$ diverges at the (outer) horizon $r=r_+$ and vanishes for $r$ approaching infinity,
 \begin{align}			 
 \lim_{r\rightarrow \infty} h|_{q=0,\theta=0} & = \lim_{r\rightarrow \infty}\left( -\frac{1}{2}\ln \left.\mathcal{P}\right|_{\theta=0}\right)=0.  
 \end{align}
 A necessary condition for the existence of polar clouds is that the first derivatives of $h$ vanish on the axis $\theta=0$. The first derivative of $T_1$ with respect to $r$ is given by 
 \begin{align}
 \partial_r\left.T_1\right|_{\theta=0}=-\frac{1}{2\left.\mathcal{P}\right|_{\theta=0}} \frac{2(r^2-a^2)}{(r^2+a^2)^2}\, , 		
 \end{align}
 which becomes zero only at $r=a\leq r_+$. This behavior of $T_1$ for $\theta=0$ is shown in figure \ref{pplot}(a). Therefore, in this case no polar clouds are possible.

  \begin{figure}
  	\includegraphics{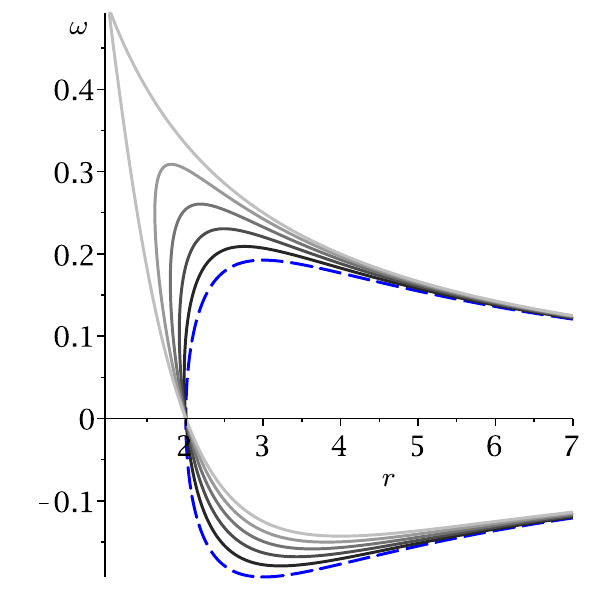} 
 	\caption{$\left.\mathcal{P}\right|_{\theta=\pi/2}=0$ as a function of $r$ and $\omega$ for different values of $a$. The blue dashed line shows the limit of $a=0$. From black to light gray the value of $a$ rises in $0.2$- steps to $a=1$. The fluid of the torus can only rotate at $r$ with an $\omega$ for which $\left.\mathcal{P}\right|_{\theta=\pi/2}>0$, which is fulfilled in the area enveloped by the graphs. The shift, especially at small radii, of allowed $\omega$ to larger values as $a$ grows is due to the frame dragging effect. }
 	\label{p0plot}
 \end{figure}
 
Now lets turn to equatorial tori ($\theta=\pi/2$). The function $\mathcal{P}$ reads
  \begin{align}
 \left.\mathcal{P}\right|_{\theta=\frac{\pi}{2}} & =\frac{1}{r}\left(2-r+4a\omega-\left(r(r^2+a^2)+2a^2\right)\omega^2 \right)\, , \label{Pg0equatorial}
 \end{align}
 which now depends on the angular velocity $\omega$. Figure \ref{p0plot} shows $\left.\mathcal{P}\right|_{\theta=\pi/2}=0$ as a function of $r$ and the angular velocity $\omega$ for different values of $a$. The condition $\mathcal{P}>0$ is satisfied between the two solutions $\omega_{1},\omega_{2}$ of $\mathcal{P}=0$ plotted in this figure. For $a=0$ the two solutions are symmetric, whereas for $a>0$ they become asymmetric showing the frame dragging effect. Here larger values of $\omega$ are favored, in particular for small radii. Note that in the ergoregion, which is given by $r_+<r<2$ for $\theta=\pi/2$, it is required that $\omega>0$. In Fig. \eqref{pplot} (b) it is shown that bigger values of $\omega$ reduce the allowed range of radii, where the effective potential is defined. 

 The effective potential does not show any local maximum for $\theta=\pi/2$. This is because the second derivative of $T_1$, 
 \begin{align}
 \partial_{rr}^2\left.T_1\right|_{\theta=\frac{\pi}{2}}=&\frac{\left(\omega(2a^2+r^3)-2a\right)^2+2r^3}{r^3\left(2a^2+r^3\right)\,\left.\mathcal{P}\right|_{\theta=\frac{\pi}{2}}}\nonumber\\
 &+\left.\frac{\left(\partial_r\mathcal{P}\right)^2}{2\mathcal{P}^2}\right|_{\theta=\frac{\pi}{2}}
 \end{align}	
 is always positive for $\left.\mathcal{P}\right|_{\theta=\frac{\pi}{2}}>0$. Therefore, equatorial tori are also not possible.
 
 Like in the Schwarzschild case no equilibrium can be found for fluid structures in rigid rotation in case of an uncharged fluid or a vanishing $B$-field.
 
 \subsection{Charged case \label{chargedcase}}
 We will now discuss some general features of the charged case, before we explicitly construct equilibrium solutions in the next sections. As we showed in the preceding subsection, a bound solution is only possible if the second term in Eqn. \eqref{potrigid} $\left(\int f_\mathcal{K}(S)\, dS\right)$ does not vanish. The function $f_\mathcal{K}(S)$ describes the charge distribution throughout the torus or polar cloud. The interaction of the fluid with the electromagnetic field results in a repulsive force in direction of $\nu$, if 
 \begin{align}
 	f_\mathcal{K}(S)\, \partial_\nu S = f_\mathcal{K}(S) (\partial_\nu A_t +\omega \partial_\nu A_\phi)>0\, .
 	\label{forcedef}
 \end{align}
 (see Eqns. \eqref{forcedir}, \eqref{kkk} and \eqref{kdef}). This force stabilizes the fluid so that equilibrium solutions can be found. The term $f_\mathcal{K}(S) \partial_\nu A_t$ in Eqn. \eqref{forcedef} corresponds to an electric field acting on a charged fluid, while the second term  $f_\mathcal{K}(S) \omega \partial_\nu A_\phi$ corresponds to the Lorentz force acting on a moving charge in a magnetic field. Both terms might independently result in an attractive or a repulsive force in direction of $\nu$, depending on the choice of $\omega$.
 
 In the following discussions we will set $f_\mathcal{K}(S)$ to 
 \begin{align}
 f_\mathcal{K}(S)=k\, S^n\, .
 \label{ffunction}
 \end{align}
 Here $k$ is a scaling factor correlated to the overall strength of the charge of the fluid and is determined according to Eqn. \eqref{fcenter} as $k=b/S^n(r_c,\theta_c)$. The exponent $n$ determines how strongly the charge distribution changes with $S$, which, in turn, changes along the fluid structure.
 
 \begin{figure}
  	\includegraphics{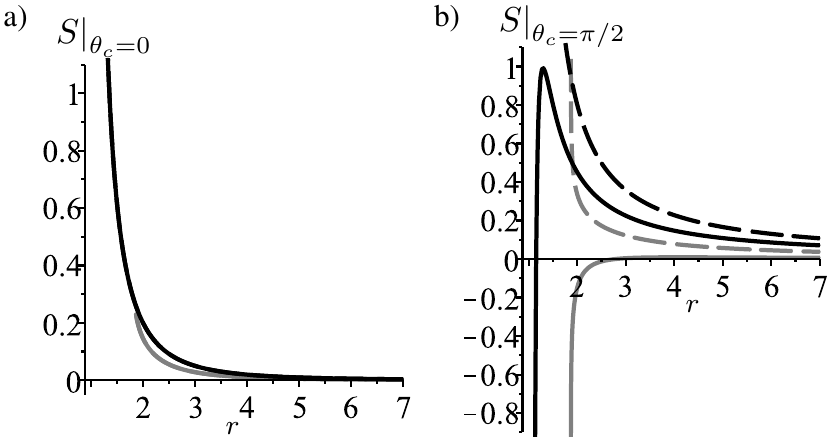} 
 	\caption{The potential $\left.S\right|_{\theta=\theta_c}$ plotted over $r$ for (a) $\theta_c=0$ and (b) $\theta_c=\pi/2$. Gray and black lines correspond to $a=1$ and $a=0.5$ respectively. Solid lines correspond to $\omega=0.04$ for $a=0.5$, and to $\omega=0.21$ for $a=1$. Dashed lines correspond to $\omega=0.12$ for $a=0.5$ and to $\omega=0.6$ for $a=1$. While $\left.S\right|_{\theta=0}$ is independent of $\omega$ and for $a<1$ approaches a finite limit at the outer horizon, $\left.S\right|_{\theta=\pi/2}$ diverges either to $+\infty$ or $\-\infty$ at the horizon, depending on the value of $\omega$ }
 	\label{Splot}
 \end{figure}
 The behavior of $S=A_t+\omega A_\phi$ is shown in Fig. \ref{Splot} on the equatorial plane $\theta=\pi/2$ and the axis $\theta=0$. From the plot and the discussion of $A_t$ and $A_\phi$ in Sec. \ref{elmpotdisc} it is clear that $\left. S \right|_{\theta=0,\frac{\pi}{2}}$ approaches zero for $r\rightarrow \infty$. On the equatorial plane $S$ diverges at the outer horizon $r=r_+$ to $\pm \infty$, depending on the choice of $\omega$,
 \begin{align}
 \lim_{r\rightarrow r_+} \left.S\right|_{\theta=\frac{\pi}{2}} = 
 \left\{ \begin{matrix} -\infty, \,  \omega < \frac{2a r_+ (r_+-1)}{(a^2+r_+^2)^2-4a^2r_+}\\ 
 +\infty, \,  \omega > \frac{2a r_+ (r_+-1)}{(a^2+r_+^2)^2-4a^2r_+}  \end{matrix} \right.		\, .
 \label{Sbehavior}
 \end{align}  
 In the $\theta=0$ case, $S$ does not depend on $\omega$ since $\left.A_\phi\right|_{\theta=0}=0$. It furthermore doesn't diverge at the horizon, but reaches the value $S(r=r_+, \theta=0)=\frac{3Ba}{4(\xi+\xi^2)}$.
 
 The effective potential has the symmetry 
 \begin{align}
 	h(a,\omega,B\,f_\mathcal{K}(S))=h(-a,-\omega,-B\,f_\mathcal{K}(S))\, .
 \end{align}
 Since $f_\mathcal{K}(S)=\frac{\rho_q}{p+\epsilon}\, U^t$, and $U^t>0$, the change $B\,f_\mathcal{K}(S)\rightarrow -B\,f_\mathcal{K}(S)$ implies either a flip of the $B$-field or a change of the fluid's charge to $-\rho_q$. The symmetry allows us to restrict the discussion to $a\geq 0$.
 
 To find tori solutions at the equatorial plane the exponent $n$ in Eqn. \eqref{ffunction} will be picked from the natural numbers ($n=0,1,2,..$), since $\left.S\right|_{\theta=\pi/2}\leq 0$ at some radii for certain $\omega$. In contrast to that, in case of polar clouds $\left.S\right|_{\theta=0}>0$ for all radii independent of $\omega$. Here $n$ can be chosen from the real numbers.

\section{Equatorial tori \label{equatorialtori}}
To construct solutions for equatorial tori, we follow the procedure introduced in Sec. \ref{procedure} and search for areas  of $\omega$ and $r_c$, where local maxima of the effective potential $h$ can be found. These areas will be influenced by the choice of the remaining parameters $n$ and $a$. $k$ in Eqn. \eqref{ffunction} is already determined by satisfying the neccessary conditions, while the values of the magnetic dipole $B$ and $\kappa$ and $\Gamma$ from the polytropic equation of state don't influence the existence conditions for a local maximum in the effective potential.

First we recall from the discussion in Sec.~\ref{procedure} that both the necessary conditions \eqref{excon1} for a maximum of the effective potential $h$ hold on the equatorial plane if we normalize the charge distribution function $f_\mathcal{K}(S)$ according to \eqref{fcenter} and \eqref{fdef}. Furthermore, the general condition $\mathcal{P}>0$ (see \eqref{Pdef} and \eqref{potrigid}) has to hold, which we already discussed in Sec. \ref{unchargedlimit}, see Eqn.~\eqref{Pg0equatorial} and Fig.~\ref{p0plot}. It therefore remains to investigate the sufficient conditions \eqref{ineqex1} and \eqref{ineqex2} for the case $\theta_c=\pi/2$. 

	\begin{figure*}
	\includegraphics{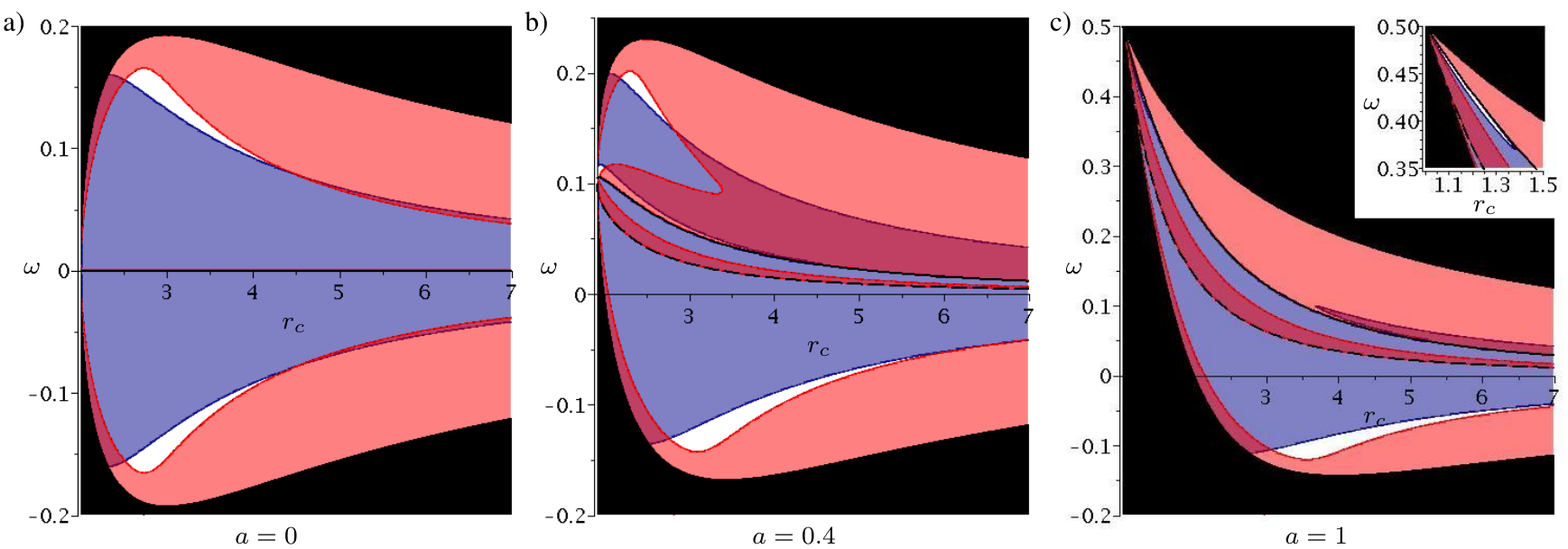}
	\caption{ Extremal points of the effective potential $h$ for $\theta_c=\pi/2$, $n=2$, and three different values of $a$ as functions of $r_c$ and $\omega$. The scaling parameter $k$ from Eqn. \eqref{ffunction} is chosen such that the conditions in \eqref{excon1} for an extremum are satisfied, hence $k$ changes throughout the plot. The extremal point corresponds to a local maximum in the white region. The value of $a$ effects the size and position of this region. Points in the light red (light gray) area correspond to a maximum in $\theta$-direction only ($\partial^2_{rr}h>0$), while points of the blue (medium gray) area correspond to a maximum only in $r$-direction ($\partial^2_{\theta\theta}h>0$).
		Points in the dark red (dark gray) area correspond to local minima in $h$. In the black area $(U^t)^2<0$, so no solutions are possible there. $\partial_r S=0$ and $S=0$ are plotted as solid and dashed black lines respectively. They mark two borders of the area, where maxima in $r$-direction are present.  $\partial_r S=0$ marks also a border of the area of maxima in $\theta$-direction. This comes due to the fact, that $\partial_r S$ and $S$ appear in the denominator in the inequalities \eqref{ineqex1}-\eqref{ineqex2}.}	
	\label{fig5}
\end{figure*}

	\begin{figure*}
	\includegraphics{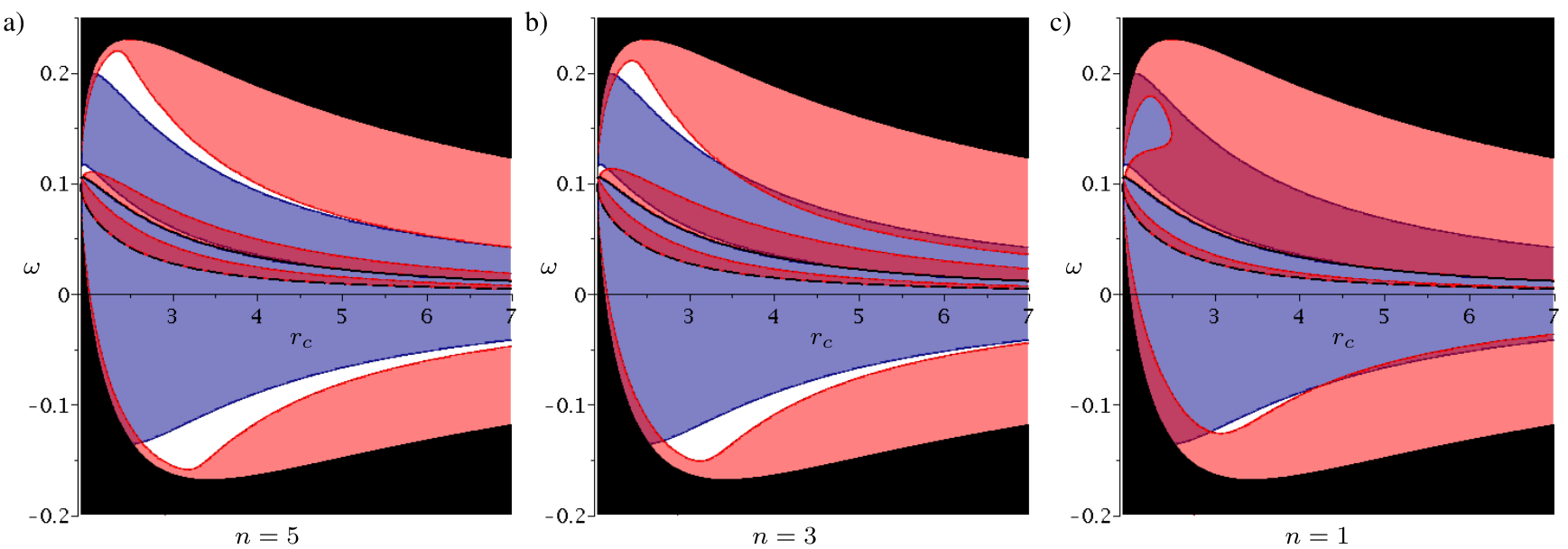}
	\caption{Extremal points of effective potential $h$ for $\theta_c=\pi/2$, $a=0.4$ and three different values of $n=5,3,1$ as a function of $r_c$ and $\omega$. The plot for $n=2$ is already presented in Fig. \ref{fig5} (b). For a detailed description see the caption of Fig. \ref{fig5}. Higher values of $n$ lead to a bigger white area, and therefore increase the parameter set of $(r_c,\omega)$ for which solutions for equatorial tori can be found. }
	\label{fig14}
\end{figure*}

Let us first discuss the influence of the rotation parameter $a$ on the existence of tori in the equatorial plane. Fig. \ref{fig5} shows the two sufficient conditions \eqref{ineqex1}, \eqref{ineqex2} together with $\mathcal{P}>0$ as functions of $r_c$ and $\omega$ for different values of $a$. The white areas indicate where all conditions are satisfied so that a maximum of the effective potential exists and a torus can be constructed. Since the scaling parameter $k$ in Eqn. \eqref{ffunction} is chosen according to Eqn. \eqref{fcenter}-\eqref{fdef}, the value of $k$ changes throughout the plot.  For the Schwarzschild case $a=0$ the plot is symmetric in the angular velocity $\omega$ and tori can be found for positive and negative values of $\omega$ quite close to the central object. For bigger $a$ however the white area moves to even smaller radii in the co-rotating case, while in the counter-rotating case it moves to bigger radii. This behavior, found for the counter-rotating case, is also seen in case of a homogeneous electromagnetic test field \cite{trovar18}. For $a=0.4$ the white area at positive $\omega$, corresponding to co-rotating tori solutions, has decreased in size. For $a=1$ the area has nearly vanished and co-rotating tori solutions can only be found at extremely small radii $r_c<1.4$ and very high angular velocities. Within this setup counter-rotating tori seem to be favored for bigger values of $a$. 
 
For all the white parameter areas in Fig. \ref{fig5}, where tori solutions can be found, $\partial_r \mathcal{P}>0$ holds, which corresponds to a repulsive force on the charged fluid in radial direction (see Sec. \ref{procedure}), which implies that the inequality \eqref{forcedef} is satisfied for $\nu=r$. While for $\omega<0$ both terms in \eqref{forcedef} correspond to a repulsive force on the torus in $r$-direction, for $\omega>0$ the electric part $\partial_r A_t$ and the magnetic part $\omega\partial_r A_\phi$ have opposite signs. Depending on the value of $\omega$ as given in \eqref{Sbehavior}, one of the two terms will dominate the divergence at $r=r_+$. The charge of the fluid or the direction of the $B$ field have then to be chosen such that the dominant term at $r\rightarrow r_+$ leads to a repulsive force on the torus. The other term, however, counteracts to the repulsive force. This leads to the reduction of the white parameter area in Fig. \ref{fig5} for bigger values for $a$ and $\omega>0$. 

\begin{figure}
		\includegraphics{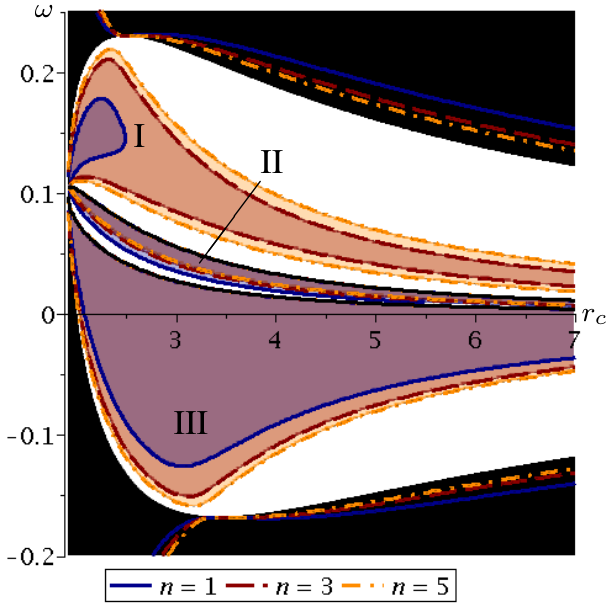}
		\caption{Extremal points of $h$ for $\theta_c=\pi/2$ and $a=0.4$. The value of $k$ changes throughout the plot, to satisfy the conditions in \eqref{excon1} for an extremum. Points in the colored area between the curves correspond to a maximum in $r$-direction.  Of the three areas (I,II,III), where maxima can be found, area I and III grow for bigger $n$, while area II shrinks for bigger $n$. In the black area $(U^t)^2<0$, so no solutions are possible there. }
		\label{fig13}
	\end{figure}

Fig. \ref{fig14} pictures the influence of the second parameter $n$ onto the size of the white parameter areas $(r_c,\omega)$, where tori solutions can be constructed. As one can see in Fig.~\ref{fig14} the white area increases for bigger $n$. The Parameter $n$ influences the area of possible solutions only by its contribution to the first sufficient condition $\partial^2_{rr}h<0$ given in \eqref{ineqex1}, where $\frac{f_\mathcal{K}'}{f_\mathcal{K}}(S)=\frac{n}{S}$ for a $f_\mathcal{K}(S)$ as given in Eqn. \eqref{ffunction}. If the prefactor of $n$ in \eqref{ineqex1} is negative, which happens if $\partial_r \mathcal{P}\, \partial_r S/S<0$, then the parameter area, where the first sufficient condition \eqref{ineqex1} holds, will increase for bigger $n$. This is also pictured in Fig. \ref{fig13}, where we directly compare the development of the areas corresponding to condition \eqref{ineqex1}, represented in Fig.~\ref{fig14} as the sum of the blue (medium gray) and white areas, for different values of $n$. At the regions I and III, that contribute to the white area in Fig.~\ref{fig14}, $\partial_r \mathcal{P}\, \partial_r S/S<0$ is satisfied and they therefore grow for bigger values of $n$. The higher the changes in the charge distribution, indicated by a bigger value of $n$, the bigger the range of parameters $(r_c,\omega)$, where solutions can be found. Even though not explicitly shown here in a plot, we want to mention, that no solutions for a bound equatorial structure can be found for $f_\mathcal{K}(S)=const.$ (corresponding to $n=0$) for $0\leq a\leq 1$.

\begin{figure}
		\includegraphics{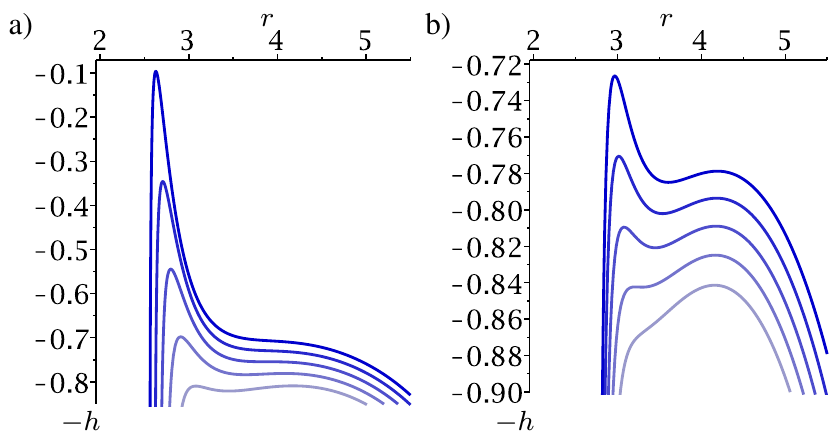}
		\caption{The negative effective potential $-h$ along the equatorial plane for $n=2, \mu=k\,(\omega B)^{n+1}=-1.929$ and different values of (a) the rotation parameter $a$ and (b) the angular velocity $\omega$. In (a) we chose $\omega=-0.1129$, and $a$ runs from $0.7$ (dark blue) to $1$ (lightest blue) in steps of $\Delta a =0.075$. In (b) we us the extremal $a=1$ and $\omega$ runs from $-0.1109$ (dark blue) to $-0.1149$ (lightest blue) in steps of $\Delta \omega=-0.001$. Due to rigid rotation the area of $r$, where a fluid torus can exist, shrinks with a growing value of $|\omega|$. The same behavior can be found in (a) for a shrinking value of $a$. For bigger $a$ and $|\omega|,\omega<0$ the torus center moves to smaller radii.}
		\label{fig6}
	\end{figure}

	For comparison with earlier related studies of charged equilibrium structures \cite{kovar14,kovar16,trovar18} we now introduce a new parameter $\mu=k\,(\omega B)^{n+1}$ used instead of the scaling factor $k$ introduced in \eqref{ffunction}. In Figs.~\ref{fig5} and \ref{fig13} we always chose $k$ such that the necessary conditions \eqref{excon1} hold, which means that $k$ changes throughout the plots. In contrast, Fig. \ref{fig6} shows the negative effective potential $-h$ along the equatorial plane for different values of $a$ and $\omega$, but for constant values of $\mu$ and exponent $n$. Due to the assumption of rigid rotation the extrema of the curve have to move closer together for bigger absolute values $|\omega|$ of the angular velocity. The same effect can be seen for changing the rotation parameter $a$ to higher values and negative $\omega$. Intuitively, this can be traced back to the frame dragging effect, due to which the same value of $\omega$ should appear smaller in a locally non-rotating reference frame in the case of bigger $a$. The torus center (appearing in Fig. \ref{fig6} as the minimum of $-h$) moves towards smaller radii for an increasing $a$ or $|\omega|$. If $a$ or $|\omega|$ are chosen too big or small for the remaining parameters of $(\mu,n,a,\omega)$ the minimum and one maximum in Fig. \ref{fig6} vanish and no bound solution can be found for the respective set of parameters. The torus solution might possess an inner cusp, through which the accretion onto the central object can occur, if the inner maximum of $-h$ has a smaller value than the outer one. In the opposite case an outer cusp might exist, where material outflow away from the central object is possible. As we can see in Fig. \ref{fig6} a slight change of $|\omega|$ or $a$ to smaller values can change the found structure from one with an inner cusp to one with an outer cusp. It might even result in the vanishing of the bound solution (see curve for $a=0.7$ in Fig. \ref{fig6} (a) or curve for $\omega=-0.1149$ ib Fig. \ref{fig6} (b)).
	
	The plotted curves for the effective potential $-h$ show the same structure as in the Schwarzschild case, discussed by \citet{kovar16}.
	
	 \begin{figure*}
    	\includegraphics{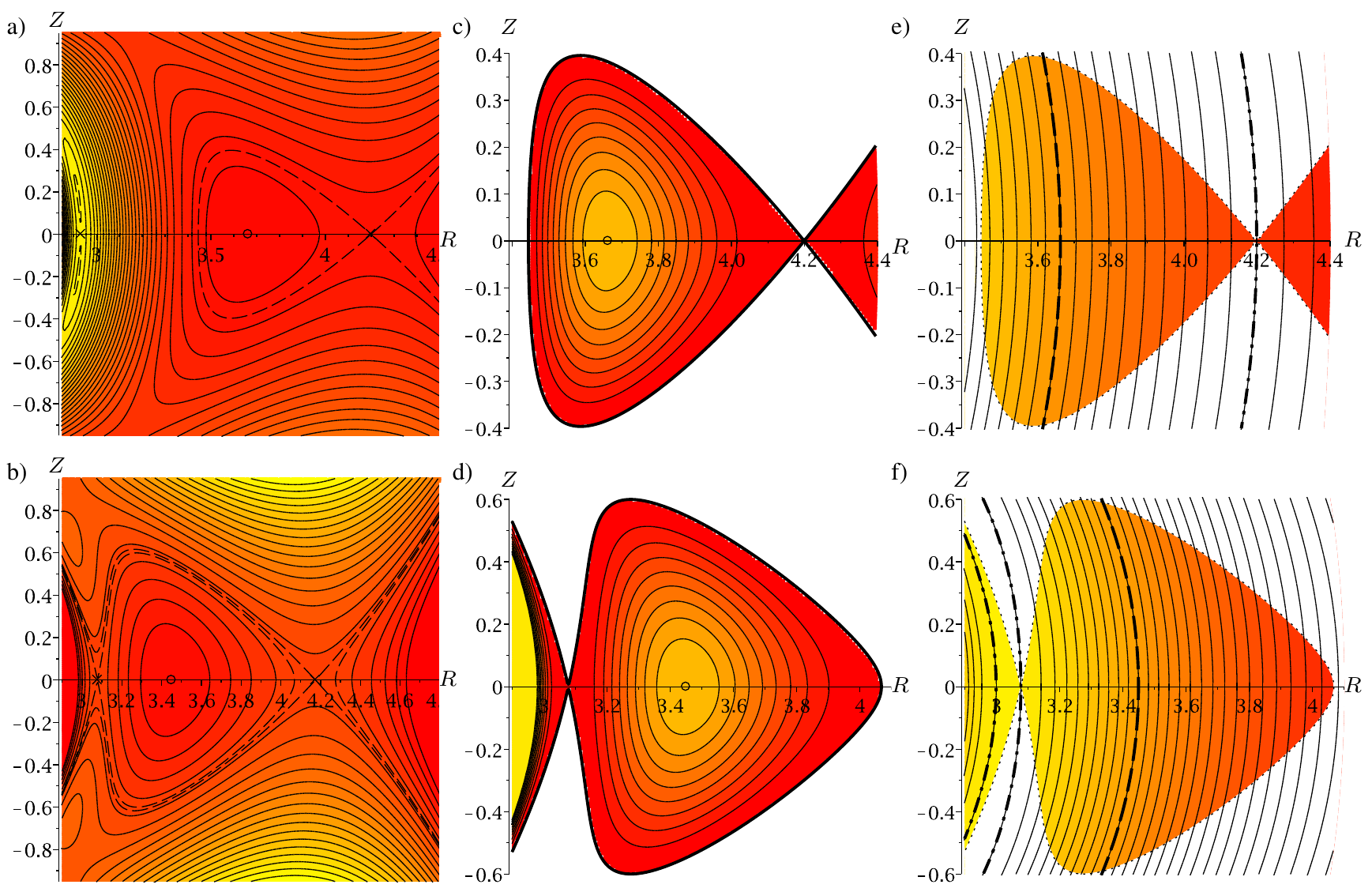}
    	\caption{Two examples A,B of equatorial tori with $n=2$, $\mu=k\,(\omega B)^{n+1}=-1.929$ and $a=1$. Solution A is presented in the first row (a,c,e), where $\omega=-0.11$. Solution B is presented in the second row (b,d,f), where $\omega=-0.1129$. The first column (a,b) shows the effective potential in the form of $-h$, the second column (c,d) the energy density distribution $\epsilon$, and third column (e,f) the specific charge distribution $q$. A red shade indicates smaller values, while a yellow shade indicates bigger values of $-h$, $\epsilon$ and $q$ respectively. The torus center is marked with a small circle in (a-d). Equipotential curves of the cusp points are plotted as dashed lines in (a,b). The energy density $\epsilon$ falls off from $\epsilon_c$ at the center to zero at the edge of the torus. Solution A shows an outer cusp, allowing matter to outflow from the torus through the cusp away from the central object. The central energy density is given by $\epsilon_c\approx8.939 \times 10^{-16}$, while the specific charge $q$ decreases towards bigger radii from $q\approx6.711 \times 10^{6}$ at the dashed line to $q\approx4.149 \times 10^{6}$ at the dash dotted line. The total charge of the torus is $\mathcal{Q}\approx4.78 \times 10^{-8}$ or $\mathcal{Q}_{SI}\approx8.197 \times 10^{12} m_n $As, where $m_n$ is the mass of the central object in solar masses.
    	Solution B shows an inner cusp, allowing matter outflow through the cusp onto the central object. The central energy density is $\epsilon_c\approx3.451 \times 10^{-15}$. The specific charge $q$ decreases towards bigger radii from $q\approx9.470 \times 10^{7}$ at the dash dotted line to $q\approx7.457 \times 10^{7}$ at the dashed line.  The total charge of the torus is $\mathcal{Q}\approx4.23 \times 10^{-7}$ or $\mathcal{Q}_{SI}\approx7.26 \times 10^{13} m_n $As.}
    	\label{fig10}
    \end{figure*}

	To make sure, that the internal magnetic field $\mathcal{B}$ of the charged torus can be neglected compared to the external magnetic test field $B$, the polytropic coefficient $\kappa$ of the equation of state \eqref{eos} will be chosen such that $\left|\frac{\mathcal{B}}{B}\right|<0.05$. To estimate the magnetic field created by the fluid structure, Eqn. \eqref{bapprox} is used. The fluid has to be diluted enough, so that the named assumption is not violated. 
	For the examples presented in Fig.~\ref{fig10} with a dipole moment $B=4.2 \times 10^{-7}$ (corresponding to $B_{\rm SI}=10^8 \, \rm T$) of the external field and a polytropic exponent $\Gamma=5/3$, the polytropic coefficient was set to $\kappa=2 \times 10^7$.

	We finally discuss two specific examples of tori in the equatorial plane. To highlight the effects of the rotation parameter $a$ as compared to the Schwarzschild case analyzed in \cite{kovar16} we choose an extremal Kerr spacetime with $a=1$. Both solutions have the same set of parameters related to the charge distribution $f_\mathcal{K}(S)$ of the torus, namely $n=2$ and $\mu= k(\omega B)^2=-1.929$, but rotate with a different angular velocity $\omega$. The equipotential surfaces, energy density and specific charge distribution are plotted for the two cases in Fig.~\ref{fig10}. For the first solution we chose $\omega=-0.11$, which then possesses an outer cusp, while the second solution with $\omega=-0.1129$ has an inner cusp. Both structures are located at rather small radii. The first torus is centered at $r_c\approx 3.66$, with a central energy density $\epsilon_c\approx 8.939 \times 10^{-16}$ and specific charge density $q_c\approx 6.711 \times 10^6$. 
	The second torus has $r_c=3.45$ (which was used to determine the value of $\mu$ applied in both cases, using Eqns. \eqref{fdef} and \eqref{ffunction}), with $\epsilon_c\approx3.451 \times 10^{-15}$ and $q_c\approx7.457 \times 10^7$. The total charge of the tori are $\mathcal{Q}=4.78 \times 10^{-8}$ and $\mathcal{Q}=4.23 \times 10^{-7}$ respectively. The specific charge distribution decreases towards bigger radii in both cases, meaning the fluid is more strongly charged closer to the central object.
	
	 The same course in the charge distribution is present in the example for rigid rotation in the Schwarzschild case. However, the example discussed by \citet{kovar16} is a very tiny structure with a diameter of $d\approx 0.02$. This structure obviously has a much smaller central density and total electric charge $\mathcal{Q}\sim 10^{-13}$. The specific charge density lies in the same order of magnitude with $q_c\approx 6.2 \times 10^6$.

\section{Polar clouds \label{polarclouds}}   
We will now discuss the construction of equilibrium structures centered on the axis $\theta=0,\pi$, termed polar clouds. It was shown in \cite{kovar16} that such structures can not exist in the Schwarzschild case. The rotation however induces an electric field on the axis $\theta=0,\pi$ given by 
\begin{align}
F_{rt} = -\frac{3 a B}{\xi^2 \Sigma}\left( \left(r^2-a^2 \right)\frac{1}{2\xi}\ln{\frac{r-1+\xi}{r-1-\xi}} -r-a^2\right),
 \label{Frt}
\end{align}
which may counteract the gravitational attraction. Note that all other components of the electromagnetic tensor vanish on the axis $\theta=0,\pi$ and, therefore, polar clouds are symmetric with respect to the equatorial plane. 

An equilibrium structure can be constructed if the effective potential $h$ has a local maximum, which happens if the necessary conditions \eqref{excon1} and the sufficient conditions \eqref{ineqex1}, \eqref{ineqex2} hold along with the general condition $\mathcal{P}>0$, see \eqref{Pdef}. As in the case of equatorial tori discussed in the forgoing section, for $\theta=0,\pi$ the necessary conditions can be fulfilled by normalizing the charge distribution function according to \eqref{fcenter} and \eqref{fdef}. The condition $\mathcal{P}>0$ reduces for $\theta=0$ to $r>r_+$, where $r_+$ is the outer horizon. Therefore, we now discuss the two sufficient conditions \eqref{ineqex1} and \eqref{ineqex2}. As the rotation of the central object is crucial for the existence of polar clouds, we focus on the influence of $a$.

\begin{figure}
		\includegraphics{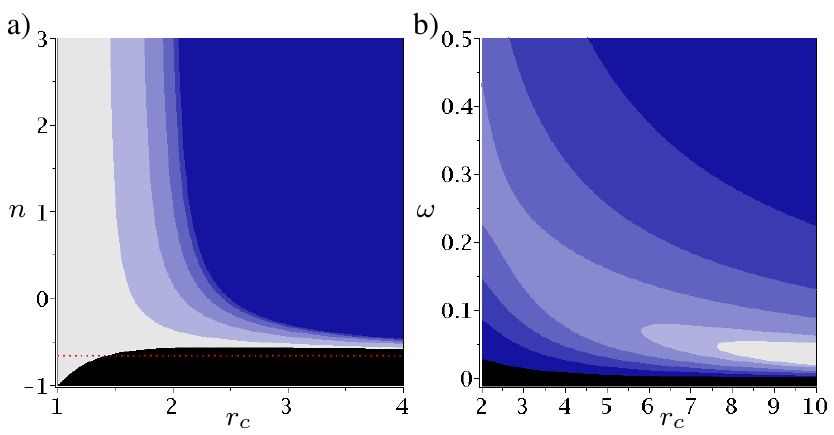}
		\caption{
		The two sufficient conditions (a) \eqref{ineqex1} and (b) \eqref{ineqex2} for $\theta_c=0$. The scaling factor $k$ changes throughout the plots to satisfy the necessary conditions \eqref{excon1} at $(\theta_c=0,r_c)$. (a) A local maximum only in $r$-direction exists for parameter sets $(r_c,n)$ from the dark blue area for $a=0.1$. The area grows for bigger $a$ (indicated by lightening up the blue color) from  $a=0.3,0.5,0.7,0.9$ to $a=1$ (white).
		For the meaning of the red dotted line at $n=-2/3$ see the text. (b) A local maximum only in $\theta$-direction occurs for parameter sets $(r_c,\omega)$ from the white area for $a=1$. The area grows for smaller $a$ from  $a=0.9,0.7,0.5,0.3$ to $a=0.1$ (blue color). For large values of $a$ the allowed values of $r_c$ are bounded from below. In the black areas in (a) and (b) the corresponding sufficient condition \eqref{ineqex1} (for (a)) or \eqref{ineqex2} (for (b)) are not fulfilled for any $0\leq a\leq 1$.}
		\label{fig7} 
	\end{figure}

The first sufficient condition \eqref{ineqex1}, which corresponds to a maximum in radial direction only, becomes independent of $\omega$ for $\theta_c=0$. It is therefore presented in Fig.~\ref{fig7}(a) as a function of $(r_c,n)$ for different values of $a$. The area of $r_c$, satisfying the condition for a maximum in $r$-direction, grows towards smaller radii for bigger values of $a$. In the limit $r_c\rightarrow \infty$ the condition \eqref{ineqex1} holds for all $n>-2/3$, while for $r_c\rightarrow r_+$ $n$ diverges to $+\infty$ for $a<1$ in order to satisfy the condition. For $a=1$ the first sufficient condition is satisfied for $n>-1$ at $r_c\rightarrow r_+$. If $a$ is not too close to $a=1$, say below $a = 0.99$, for $n<-2/3$ no maximum exists for any value of $r_c$.
	
The second sufficient condition \eqref{ineqex2}, which corresponds to a maximum in $\theta$-direction only, is presented in Fig. \ref{fig7} (b) for the parameter space $(r_c,\omega)$ and different values of $a$. This condition is independent of the parameter $n$, and can only hold for co-rotating clouds $\omega>0$. 

An attractive force towards the rotation axis, produced by the Lorentz-force on the rotating charged fluid ($f_\mathcal{K}(S)\partial_\theta S<0$ for some area $0<\theta<\Delta \theta$), is necessary to find a local maximum of $h$ at the poles. A local maximum further requires a repulsive force in $r$-direction, which is created on the polar axis solely by the local electric field component, arising from $A_t$, and acting on the charged fluid. $A_\phi$ and its derivatives vanish on the polar axis. This requirement determines how the torus is charged $(\left.f_\mathcal{K}(S)\partial_r A_t\right|_{\theta=0}>0)$. Since $\partial_r A_t(r,\theta=0)<0$, $f_\mathcal{K}(S)$ has to be negativ on the polar axis as well as in some area $0<\theta<\beta$, for which $S(r,\theta)$ does not change its sign. An attractive force can now only be achieved close to the polar axis, if $\partial_\theta S=\partial_\theta A_t+\omega \partial_\theta A_\phi>0$. Since $\partial_\theta A_t<0$ and $\partial_\theta A_\phi>0$ for $0<\theta<\pi/2$ and $r>r_+$, this condition can only be satisfied in the co-rotating case $\omega>0$. This result coincides with the one found by \citet{trovar18} for a homogeneous magnetic test field and a central object without a net charge ($e=0$ in their notation). 

The area, where condition \eqref{ineqex2} holds, i.e.~where maxima in $\theta$-direction exist, is largest for small $a>0$. However, keep in mind, that the scaling factor $k$ in the overall charge distribution of the fluid, given by $f_{\mathcal{K}}(S)$, changes throughout the plot to satisfy the necessary condition in \eqref{excon1} for an extremal point at $(\theta_c=0,r_c)$. According to Eqn. \eqref{fcenter} and \eqref{ffunction}, $k$ diverges for $\partial_r S\rightarrow 0$, which is the case for $r_c\rightarrow \infty$ or $a\rightarrow 0$. For small $a$, solutions can be found for a wide set of parameters $(r_c,\omega)$, however a strongly charged fluid is required in this case. 

\begin{figure}
		\includegraphics{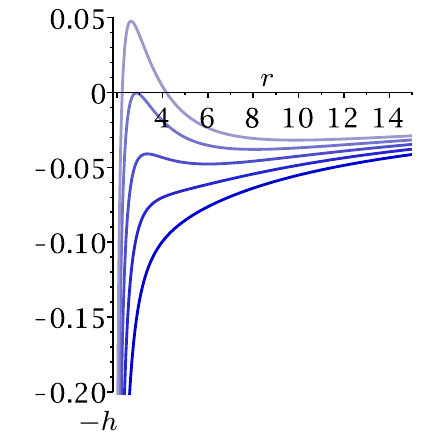}
		\caption{Different plots of the negative effective potential $-h$ along the rotation axis $(\theta=0)$ for $n=-1/2, \mu=k\,(\omega B)^{n+1}=-0.596$ and different values of $a$. The effective potential $h$ is independent of $\omega$ for $\theta=0$. The rotation parameter $a$ runs from $0.2$ (dark blue) to $0.4$ (lightest blue) in steps of $\Delta a =0.05$.  For bigger $a$ the torus center moves to bigger radii.}
		\label{fig8}
	\end{figure}

Fig. \ref{fig8} shows the negative effective potential $-h$ along the rotation axis $\theta=0,\pi$ for different values of $a$, but for constant values of $\mu= k(\omega B)^n$ and exponent $n$. The repulsive effect of the electric field component $F_{rt}$, given in \eqref{Frt}, grows for bigger values of $a$, as it is the result of the frame dragging effect. This effect manifests in the plot as the growing maximum of $-h$ for bigger $a$. The center of the polar cloud, which corresponds to a minimum in Fig. \ref{fig8}, moves towards bigger radii for an increasing $a$. It can be seen in Fig.~\ref{fig8} that a minimum only exists for very specific values of $a$, and that it  vanishes and no bound solution for a polar cloud can be found if $a$ is not chosen appropriately for the respective set of parameters. The polar cloud solution might possess an inner cusp on the rotation axis, if the inner maximum of $-h$ is smaller than zero, $\left.-h\right|_{r=r_c}<0$. An outer cusp can not exist, since the effective potential does not diverge at any $r>r_c$, but approaches zero for $\theta=0$ (see Sec. \ref{unchargedlimit}-\ref{chargedcase}). Outflows from the polar cloud away from the central object, might still occur at cusp points located at $\theta\neq 0$.

\begin{figure}
		\includegraphics{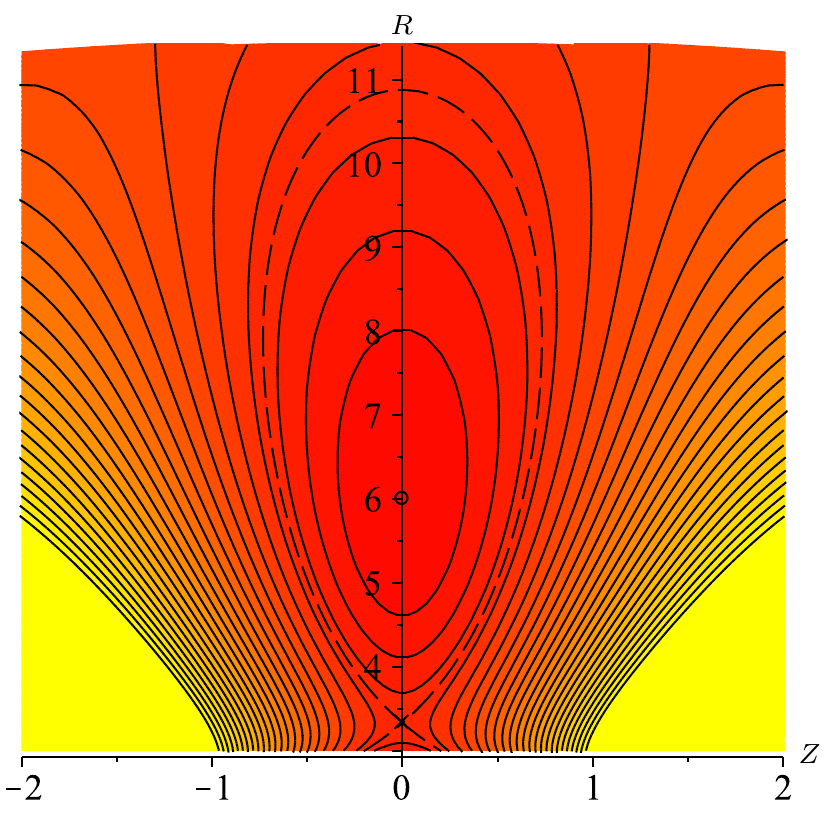}
		\caption{Effective potential in form of $-h$ of a polar cloud solution for $n=-0.5$, $\mu=k\,(\omega B)^{n+1}\approx-0.596$, $a=0.3$, and $\omega=0.1$. A red shade indicates smaller values, while a yellow shade indicates bigger values of $-h$. The center of the cloud at $r_c=6$ is marked with a small circle, the equipotential curve of the cusp point is plotted as a dashed line. The solution shows an inner cusp, allowing matter outflow through the cusp onto the central object.} 
		\label{polplot}
    \end{figure}

\begin{figure}
		\includegraphics{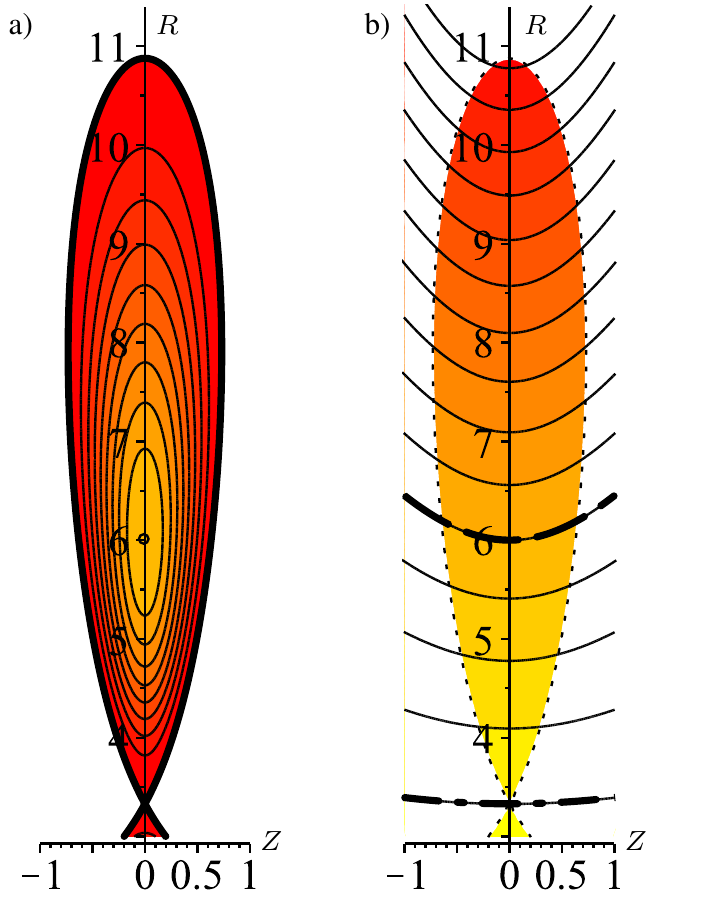}
		\caption{(a) Energy density distribution $\epsilon$ and (b) specific charge distribution $q$ of a polar cloud solution (corresponding potential shown in Fig.~\ref{polplot}) for $n=-0.5$, $\mu=k\,(\omega B)^{n+1}\approx-0.596$, $a=0.3$, and $\omega=0.1$. The center of the cloud at $r_c=6$ is marked with a small circle in (a). A red shade indicates smaller values, while a yellow shade indicates bigger values of $\epsilon$ and $q$ respectively. The energy density $\epsilon$ falls off from $\epsilon_c$ at the center to zero at the edge of the cloud. Through the  inner cusp at $r\approx 3.5$ matter can be accreted by the central object. The polar cloud has a central energy density of $\epsilon_c\approx1.5832\times 10^{-15}$. The absolute value of the specific charge $|q|$ increases towards bigger radii from $q\approx-1.040\times 10^7$ at the dash dotted line to $q\approx-1.730\times 10^7$ at the dashed line. This behavior of the charge distribution is opposite to the ones found for the examples of equatorial tori shown in Fig. \ref{fig10}, where $|q|$ decreases towards bigger radii. The total charge of the polar cloud is $\mathcal{Q}\approx9.80\times 10^{-7}$ or  $\mathcal{Q}_{\rm SI}\approx1.68\times 10^{14} m_n \rm As$, where $m_n$ is the mass of the central object in solar masses.} 
		\label{fig12}
    \end{figure}

Finally we construct a specific example of a polar cloud, using the procedure introduced in Sec. \ref{procedure}. The equipotential surfaces of a polar cloud with an inner cusp for $a=0.3$,  $n=-0.5$, $\omega=0.1$ and $r_c=6$ are plotted in Fig. \ref{polplot}. The corresponding energy density and specific charge distribution are plotted in Fig. \ref{fig12}, where, like in the case of equatorial tori, $\Gamma=5/3$, and $\kappa=2 \times 10^7$ was used for an external field with a dipole moment $B=4.2\times10^{-7}$. Again, the condition $\left|\frac{\mathcal{B}}{B}\right|<0.05$ is satisfied, so that the magnetic field produced by the polar cloud can be neglected compared to the external field. At the center the polar cloud has a central energy density of $\epsilon_c\approx1.5832\times 10^{-15}$, and a specific charge density of $q_c\approx-1.730\times 10^{7}$. The polar cloud has a total charge of $\mathcal{Q}\approx9.80\times10^{-7}$.
 The absolute value of the specific charge distribution $|q|$ increases towards bigger radii. 

\section{Summary and Conclusion \label{conclusion}}
	In this work, we studied the existence of stationary charged fluid structures around a central object with an electromagnetic test field, that does not contribute to the spacetime. We assumed that both the spacetime and the electromagnetic test field are stationary, axially symmetric, and mirror symmetric with respect to the equatorial plane. The fluid is assumed to move in this background without influencing it, which implies that it has a small charge and mass as compared to the central object and the electromagnetic test field. 
	We further assume a perfect fluid with a polytropic equation of state and zero conductivity, with spatial motion in azimuthal direction only. To satisfy the resulting integrability condition, we required a constant angular velocity throughout the fluid structure, i.e.~rigid rotation, and a charge distribution in the fluid that is given by a function of the potential of the electromagnetic test field. In this work we focus our attention  on stationary fluid structures centered on the equatorial plane, named equatorial tori, and on structures centered on the axis of symmetry, named polar clouds. The procedure described here can then be used to construct fluid structures for any spacetime and electromagnetic test field, that satisfy the named conditions.
	
	In the second part we specify the discussion to the case of a Kerr spacetime with a dipole magnetic test field, which is a direct generalization of the Schwarzschild case discussed by \citet{kovar16}. It could describe an idealized, rather compact rotating neutron star, which produces a dipole magnetic field, that is oriented along the rotation axis.
	
	In this scenario we then studied in detail the existence conditions for equatorial tori and polar clouds. We confirmed that in the uncharged limit, meaning an uncharged fluid or a vanishing test electromagnetic field, both types of structures can not exist for a rigidly rotating fluid. For the general charged case we found that the rotation of the central object has a major impact on the region of existence of stationary structures in parameter space. This can be traced back to the interplay of the electromagnetic test field and the frame dragging, which induces a time-like component in the potential associated with an electric part of the magnetic dipole field. In the case of equatorial tori this causes for high spins of the central object a preference for counter rotating tori in the equatorial plane. For polar clouds the rotation is even more essential, as this kind of structure can not exist at all in the non rotating case. As the magnetic field vanishes along the symmetry axis, the electric field is the only part which can balance the gravitational attraction. We found that for a small rotation of the compact object polar clouds can exist for a wide range of central radii $r_c$, which seems to be counterintuitive at first glance in view of the non rotating limit. However, for small values of the rotation the electric field is weak, as expected, which needs to be compensated by an extremely high charge of the fluid, which is physically unrealistic and may also violate the assumptions within our model. Furthermore, both in the case of counter rotating equatorial tori, and polar clouds the center of solutions are found farer away from the black hole for higher spins of the central object.
	
	We also discussed fluid structures which allow an outflow towards or away from the central object, encoded by the existence of cusps in the effective potential. Tori and polar clouds that possess an inner or outer cusp (the latter occurring only in the tori case) can be found by slightly varying the set of parameters like the angular velocity of the fluid, the spin of the central object, or a parameter introduced by the choice of the function that is connected to the charge distribution (corresponding to $n$ in the discussed case). We explicitly constructed examples with inner and outer cusps and discussed their physical characteristics.
	
	An open question is the choice of the function, depending on the electromagnetic potential only, that is connected to the charge distribution within the torus. For the simplest appoach- setting the funciton to constant- no solutions for equatorial tori can be found for $0\leq a\leq 1$.  Are there restriction to the free choice of the function, so that the torus solution is stable? In general it should also be possible, to choose the function such that the total charge of the torus vanishes. This is however not a straight forward task. Including selffields - let it be it gravitational or electromagnetic- could bring the model closer to the description of realistic accretion discs. It would also be very interesting to consider a fluid with nonzero conductivity. However, in this case radial motion within the fluid is to be expected, which can maybe be handled perturbatively.

\begin{acknowledgments}
The authors gratefully acknowledge support from the Research training Group 1620 ``Models of Gravity'' funded by the German Research Foundation DFG. K.S., E.H., and C.L. further thank the DFG funded Collaborative Research Center 1128 ``Relativistic Geodesy and Gravimetry with Quantum Sensors (geo-Q)'' for support. We furthermore thank P. I. Jefremov and V. Witzany for insightful discussions.
\end{acknowledgments}

\bibliography{bibliography}

\end{document}